\documentclass[a4paper,11pt]{article}
\usepackage{jheppub}
\usepackage{lineno}
\usepackage{dcolumn}
\usepackage{color}
\usepackage{graphicx}
\usepackage{adjustbox}
\usepackage[colorlinks=true]{hyperref}
\usepackage{amsmath}
\usepackage{multirow}
\usepackage{bm}
\usepackage{float}
\usepackage{caption}
\usepackage{soul}

\title{\boldmath {Measurement of two-neutrino double electron capture half-life of $^{124}$Xe with PandaX-4T}}

\collaborationImg{\includegraphics[width=0.3\textwidth]{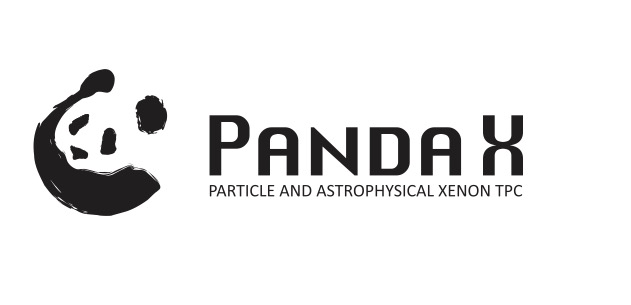}}

\collaboration{The PandaX-4T Collaboration}

\affiliation[a]{New Cornerstone Science Laboratory, Tsung-Dao Lee Institute, Shanghai Jiao Tong University, Shanghai 201210, China}
\affiliation[b]{School of Physics and Astronomy, Shanghai Jiao Tong University, Key Laboratory for Particle Astrophysics and Cosmology (MoE), Shanghai Key Laboratory for Particle Physics and Cosmology, Shanghai 200240, China}
\affiliation[c]{Shanghai Jiao Tong University Sichuan Research Institute, Chengdu 610213, China}
\affiliation[d]{Jinping Deep Underground Frontier Science and Dark Matter Key Laboratory of Sichuan Province}
\affiliation[e]{Yalong River Hydropower Development Company, Ltd., 288 Shuanglin Road, Chengdu 610051, China}
\affiliation[f]{Sino-French Institute of Nuclear Engineering and Technology, Sun Yat-Sen University, Zhuhai, 519082, China}
\affiliation[g]{Department of Physics, Yantai University, Yantai 264005, China}
\affiliation[h]{Key Laboratory of Nuclear Physics and Ion-beam Application (MOE), Institute of Modern Physics, Fudan University, Shanghai 200433, China}
\affiliation[i]{School of Physics, Beihang University, Beijing 102206, China}
\affiliation[j]{Peng Huanwu Collaborative Center for Research and Education, Beihang University, Beijing 100191, China}
\affiliation[k]{International Research Center for Nuclei and Particles in the Cosmos \& Beijing Key Laboratory of Advanced Nuclear Materials and Physics, Beihang University, Beijing 100191, China}
\affiliation[l]{Southern Center for Nuclear-Science Theory (SCNT), Institute of Modern Physics, Chinese Academy of Sciences, Huizhou 516000, China}
\affiliation[m]{SJTU Paris Elite Institute of Technology, Shanghai Jiao Tong University, Shanghai, 200240, China}
\affiliation[n]{School of Mechanical Engineering, Shanghai Jiao Tong University, Shanghai 200240, China}
\affiliation[o]{Department of Physics, University of Maryland, College Park, Maryland 20742, USA}
\affiliation[p]{School of Physics, Nankai University, Tianjin 300071, China}
\affiliation[q]{School of Physics, Sun Yat-Sen University, Guangzhou 510275, China}
\affiliation[r]{State Key Laboratory of Particle Detection and Electronics, University of Science and Technology of China, Hefei 230026, China}
\affiliation[s]{Department of Modern Physics, University of Science and Technology of China, Hefei 230026, China}
\affiliation[t]{Research Center for Particle Science and Technology, Institute of Frontier and Interdisciplinary Science, Shandong University, Qingdao 266237, Shandong, China}
\affiliation[u]{Key Laboratory of Particle Physics and Particle Irradiation of Ministry of Education, Shandong University, Qingdao 266237, Shandong, China}
\affiliation[v]{School of Physics, Peking University, Beijing 100871, China}
\affiliation[w]{College of Nuclear Technology and Automation Engineering, Chengdu University of Technology, Chengdu 610059, China}
\affiliation[x]{School of Physics and Astronomy, Sun Yat-Sen University, Zhuhai 519082, China}

\author[b]{Zihao Bo,}
\author[b]{Wei Chen,}
\author[a,b,c,d]{Xun Chen,}
\author[e,d]{Yunhua Chen,}
\author[f]{Zhaokan Cheng,}
\author[a]{Xiangyi Cui,}
\author[g]{Yingjie Fan,}
\author[h]{Deqing Fang,}
\author[b]{Zhixing Gao,}
\author[i,j,k,l]{Lisheng Geng,}
% ...

\emailAdd{ke.han@sjtu.edu.cn}
\emailAdd{jianglai.liu@sjtu.edu.cn}
\emailAdd{wangqiuhong@fudan.edu.cn}

\abstract{
Detailed studies of two-neutrino double electron capture (2$\nu$DEC) is a crucial step towards searching for the neutrinoless mode to explore the Majorana nature of neutrinos.
We have measured precisely the half-life of the 2$\nu$DEC process in $^{124}$Xe, utilizing a total exposure of 1.73 tonne$\cdot$year from the commissioning run and the first science run of the PandaX-4T experiment.
A time-dependent background model in the $\mathcal{O}$(10 keV) energy is constructed for the first time in PandaX-4T data.
With an unbinned maximum likelihood fit, we determine the half-life of the 2$\nu$DEC process to be $(1.03\pm0.15_{\rm stat}\pm0.08_{\rm sys})\times 10^{22}$\,yr.
Furthermore, we have evaluated the capture fraction for both electrons captured from the $K$ shell ($KK$) to be $(65\pm5)\%$, which aligns with the $^{124}$Xe nuclear model calculations within 1.8\,$\sigma$. 
}

\begin{document}

\maketitle
\flushbottom

\section{Introduction}
Two-neutrino double electron capture (2$\nu$DEC) is a rare nuclear process in which two protons in a nucleus simultaneously capture two orbital electrons and are converted to neutrons.
Within the framework of the Standard Model (SM) of particle physics, two neutrinos are emitted in this process.
The inverse half-life of this process, denoted by $(T_{2\nu})^{-1}$, can be expressed as
\begin{equation}\label{eq:halflife_2nu}
 (T_{2\nu})^{-1}=G_{2\nu}\left|M_{2\nu}\right|^2,
\end{equation}
where $G_{2\nu}$ is the phase space factor, and $M_{2\nu}$ is the nuclear matrix element (NME) for the process~\cite{Suhonen:1998ck}.
If neutrinos are their own antiparticles, i.e., Majorana fermions, the process above could also have a neutrinoless mode~\cite{Avignone:2007fu}, namely neutrinoless double electron capture (0$\nu$DEC).
Once this phenomenon is observed, it will confirm the Majorana nature of neutrinos, directly violate the lepton number conservation, and open new windows to physics beyond the SM~\cite{Dolinski:2019nrj, Mohapatra:1986su, Deppisch:2012nb}.
The inverse half-life of this neutrinoless process, $(T_{0\nu})^{-1}$, is given by
\begin{equation}
 (T_{0\nu})^{-1}=G_{0\nu}\left|M_{0\nu}\right|^2\left|\frac{\langle m_{\nu}\rangle}{m_{e}}\right|^{2},
\end{equation}
where $G_{0\nu}$ and $M_{0\nu}$ represent the phase space factor and NME for the neutrinoless mode, respectively, $m_e$ is the electron mass, and $\langle m_{\nu}\rangle$ is the effective Majorana mass for the electron neutrino~\cite{Belli:2021pzu, Wittweg:2020fak}.
Although $M_{2\nu}$ and $M_{0\nu}$ are not directly related, precise measurements of the half-life of 2$\nu$DEC can serve as a reference point for various NME calculation methods for both the two-neutrino and neutrino-less modes~\cite{Belli:2021pzu}.

$^{124}$Xe is an excellent candidate for 2$\nu$DEC because the phase space factor $G_{2\nu}$ is proportional to the fifth power of the reaction's relatively high Q-value of 2856~keV~\cite{Kotila:2013gea, Nesterenko:2012xp}, 
The 2$\nu$DEC process of $^{124}$Xe proceeds as
\begin{equation}\label{eq:DEC}
    \rm ^{124}Xe + 2e^- \xrightarrow {}{} ^{124}Te + 2\nu_e + X.
\end{equation}
After capturing the two electrons, the resulting two vacancies of the $^{124}$Te daughter atom are subsequently filled by emitting X-rays and/or Auger electrons, which we denote as X.
The energy deposited by these emissions is on the order of 10\,keV, a range detectable by large xenon detectors such as PandaX-4T~\cite{Zhang:2018xdp}, initially designed for dark matter searches. 
To date, the half-life of 2$\nu$DEC in $^{124}$Xe has been measured by the XENON~\cite{XENON:2019dti,XENON:2022evz,XENONCollaboration:2022kmb} and LZ collaborations~\cite{Aalbers:2024xwo}.
Besides $^{124}$Xe, other isotopes that have been observed to undergo 2$\nu$DEC include $^{78}$Kr~\cite{Gavrilyuk:2013yqa,Ratkevich:2017kaz} and $^{130}$Ba~\cite{PUJOL20096834,Meshik:2001ra}.

In this work, we report a measurement of the $^{124}$Xe 2$\nu$DEC half-life, along with the relative capture fractions for different atomic shell capture modes, based on combined data from the commissioning run (Run0) and the first science run (Run1) of the PandaX-4T experiment.
Sec.~\ref{sec.Detector} provides a brief overview of the PandaX-4T detector and its data-taking history.
Sec.~\ref{sec.Analysis} details the data analysis, including event selection, energy reconstruction, signal and background models, and the fit procedure.
The final results and comprehensive uncertainty analysis are presented and discussed in Sec.~\ref{sec.Result}.

\section{PandaX-4T detector and data-taking campaigns} \label{sec.Detector}
PandaX-4T is a multi-purpose experiment located in the B2 hall of the China Jinping Underground Laboratory (CJPL-II)~\cite{Li:2014rca}.
It uses natural xenon as the target material to search for dark matter signals and investigate the fundamental properties of neutrinos.
The detector holds 5.6 tonnes of liquid xenon and is housed within a stainless steel water tank containing 900 tonnes of ultra-pure water for shielding against external radioactivity.
The xenon cooling and purification system consists of three cooling units and two separate re-circulation loops, which continuously eliminate contaminants through hot getters~\cite{Zhao:2020vxh}.
The cylindrical dual-phase time projection chamber (TPC) has a sensitive volume with 3.7 tonnes of liquid xenon in the electric field cage.
A vertical electric field is established by the anode mesh, gate mesh, and cathode grid from top to bottom. 
The separations of neighboring electrodes are 10\,mm and 1185\,mm, respectively.
The side of the electric field cage is enclosed by 24 highly reflective polytetrafluoroethylene (PTFE) panels, and the distance between opposite panels is 1185\,mm.
The TPC is equipped with 169 and 199 Hamamatsu R11410-23 3-inch photomultiplier tubes (PMTs) at the top and bottom, respectively, which detect prompt scintillation ($S1$) and delayed electroluminescence ($S2$) photons resulting from energy depositions.
These two signals are used to precisely reconstruct both the energy and the three-dimensional position of events~\cite{PandaX-4T:2021bab, PANDA-X:2021jua, PandaX:2024med}.
The skin region between the TPC and the inner cryostat vessel serves as a veto system with 105 Hamamatsu R8520 1-inch PMTs instrumented on the top and bottom.

\begin{figure*}[tb]
  \centering
  \includegraphics[width=\columnwidth]{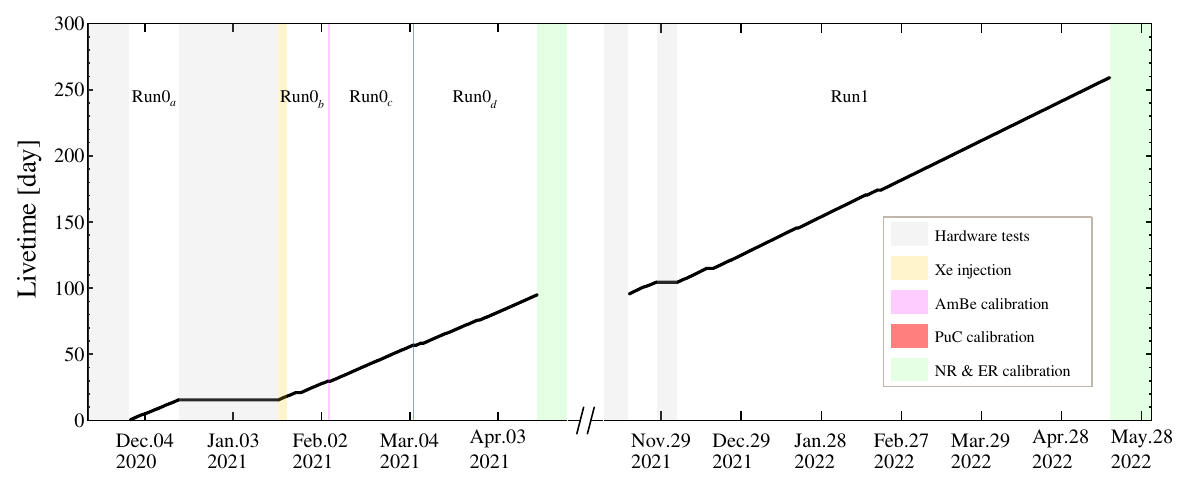}
  \caption{Data taking history and accumulated live time in Run0 and Run1. Gray shaded regions indicate the periods for hardware tests, and that for detector calibration after physics data taking is indicated by green shaded regions. Several important operation and calibration campaigns within Run0 are also marked, including xenon injection (yellow), AmBe calibration (magenta), and PuC calibration (red), which divides Run0 into four periods: Run0$_{a}-$Run0$_{d}$.
 }
  \label{fig:exposure_operation}
\end{figure*}

The data-taking and operational history of PandaX-4T is summarized in Figure~\ref{fig:exposure_operation} and is also detailed in Ref.~\cite{PandaX-4T:2021bab, PandaX:2024qfu}.
The experiment has completed two stable data-taking periods, Run0 and Run1.
Run0 took place from November 28, 2020, to April 16, 2021, with 94.9 days of physics data collected.
Following this, calibration and an offline distillation campaign aimed at tritium removal were conducted~\cite{Cui:2020bwf, Cui:2024ltd}.
Run1 resumed data-taking from November 16, 2021, to May 15, 2022, yielding 163.6 days of physics data.
Several operational and calibration campaigns, especially the xenon injection and neutron calibration, significantly impacted the background levels in this analysis and are therefore highlighted in Figure~\ref{fig:exposure_operation}, with their effects discussed in detail in Sec.~\ref{subsec.bkg}.

\section{Data analysis}\label{sec.Analysis}
\subsection{Data selection and data quality cuts}\label{subsec.dataselectQC}
The combined data from Run0 and Run1 are utilized in this analysis.
The basic data processing and signal reconstruction procedures and the data selection criteria are inherited from the dark matter search analysis~\cite{PandaX:2024med}.
Only the energy region of interest (ROI) and data quality cuts are updated specifically for this analysis.

The energy range between 25\,keV and 75\,keV is defined as the ROI in this analysis,
and the specifics of energy reconstruction are provided in Sec.~\ref{subsec.energyrecon}.
This range captures most of the energy peaks from the 2$\nu$DEC of $^{124}$Xe while minimizing the impact of background components such as $^{133}$Xe (see Sec.~\ref{subsec.sig} and Sec.~\ref{subsubsec.activation} for details).
Unlike the possible $\mathcal{O}$(keV) signals in dark matter searches, the signals in our ROI are sufficiently large. 
They are not significantly affected by isolated $S1$ signals or the \emph{afterglow} following high-energy events~\cite{PandaX:2023sil}.
Consequently, we can recover periods excluded in Ref.~\cite{PandaX:2024qfu}, and the live times for both Run0 and Run1 are equal to their calendar duration.
The same fiducial volumes (FVs) as in the dark matter search analysis are used, corresponding to fiducial masses of $2.38\pm0.04$ tonnes in Run0 and $2.48\pm0.05$ tonnes in Run1, resulting in a total exposure of 1.73 tonne$\cdot$year.

Due to the low rate of non-physical noise signals within our ROI, we have relaxed the requirement of correlation between $S2$ width and drift time in Ref.~\cite{PandaX:2024med}, which is caused by the electron diffusion effect during drifting.
For example, in the middle of the detector, the lower (upper) limit of the $S2$ width for 50\,keV events is relaxed from 2.5 (3.7)\,$\mu$s to 2.0 (4.2)\,$\mu$s.
Additionally, the charge distribution cut and the top-bottom partition cut for the $S2$ signal have been removed, as the pulse shape of the $S2$ within the ROI is already satisfactory.
By easing these cuts, we have achieved higher signal efficiency.
All other cuts are consistent with those in Ref.~\cite{PandaX:2024med}, with minor adjustments to extend from lower energy regions to the higher energy ROI.

\begin{figure}[!tb]
  \centering
  \includegraphics[width=0.7\columnwidth]{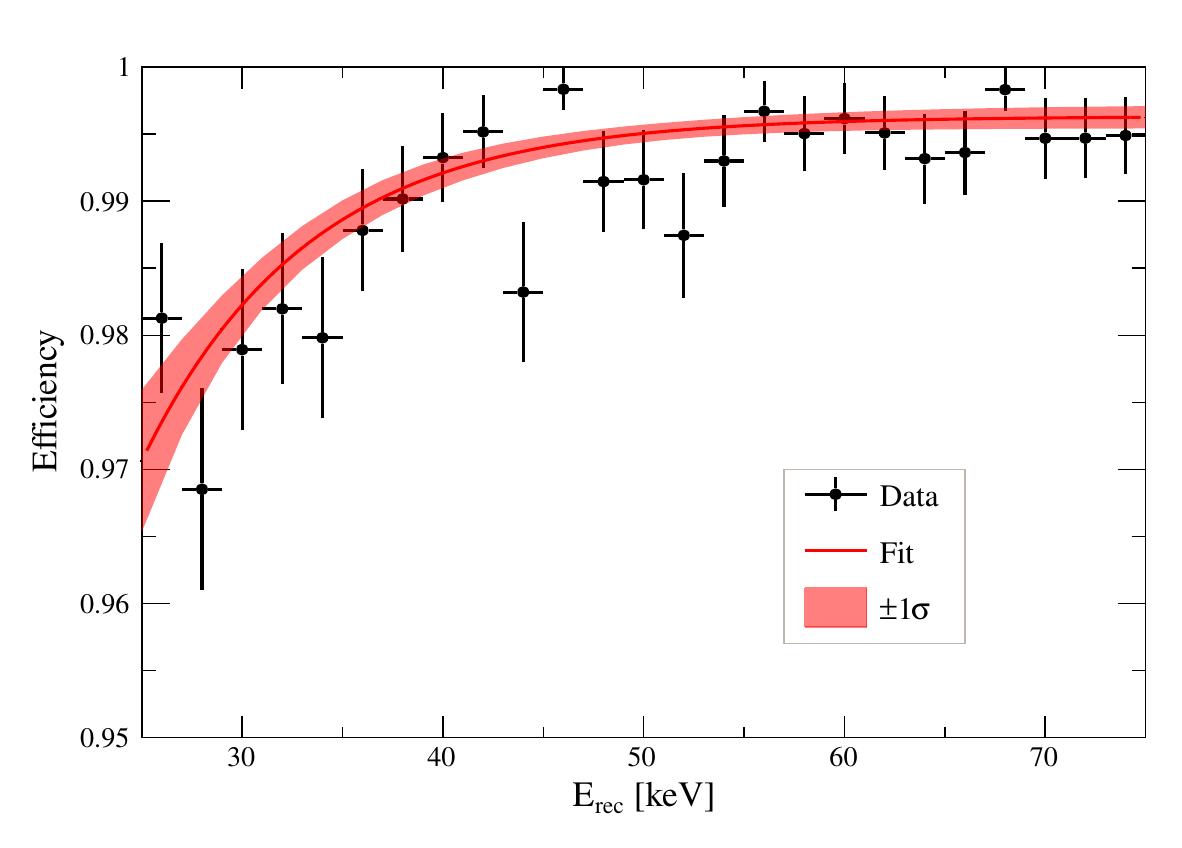}
  \caption{The data quality cut efficiency derived from $^{220}$Rn calibration data (black dots). The red curve represents the exponential fit to the data, and the red shaded band represents the associated uncertainty.}
  \label{fig:effcurve}
\end{figure}

The data quality criteria described above result in a stable efficiency of approximately 99.5\% for energies above 50\,keV, with a slight drop in the lower energy range, as illustrated in Figure~\ref{fig:effcurve}.
The efficiency is determined using $^{220}$Rn and $^{222}$Rn calibration data,
and the fitted efficiency curve with associated uncertainty is then applied to the signal and background spectra in the final fit.

\subsection{Energy reconstruction}\label{subsec.energyrecon}
Bottom-only S2 is used for energy reconstruction to avoid the effect of dead channels and saturation of the top PMTs.
The energy of a given event is reconstructed as
\begin{equation}
 E_{\rm rec} = W \times (\frac{Q^{c}_{\rm S1}}{g1} + \frac{Q^{c}_{\rm S2_{\rm b}}}{g2_{\rm b}}),
\end{equation}
where $W$ = 13.7~eV is the average energy to generate quanta in liquid xenon ~\cite{Szydagis:2011tk}. 
${Q^{c}_{\rm S2_{\rm b}}}$ and $Q^{c}_{\rm S1}$ are the $S1$ and $S2_{\rm b}$ charges that have been corrected for spatial uniformity and temporal stability~\cite{PandaX:2024med}.
The reconstruction parameters, ($g1$, $g2_{\rm b}$), are calibrated using the mono-energetic electron-recoil (ER) peaks including 41.5\,keV ($^{\rm 83m}$Kr), 164\,keV ($^{\rm 131m}$Xe), and 236\,keV ($^{\rm 129m}$Xe\,+\,$^{\rm 127}$Xe).
Unlike the dark matter analysis in the low energy range, the values of ($g1$, $g2_{\rm b}$) are separately fitted for Run0 and Run1 using the Doke-plot method~\cite{Doke:2002oab} to optimize energy reconstruction accuracy within our ROI, yielding $(0.100 \pm 0.001, 4.0 \pm 0.1)$ for Run0 and $(0.096 \pm 0.001, 4.5 \pm 0.1)$ for Run1.

\begin{figure}[!hbtp]
  \centering
  \includegraphics[width=0.7\columnwidth]{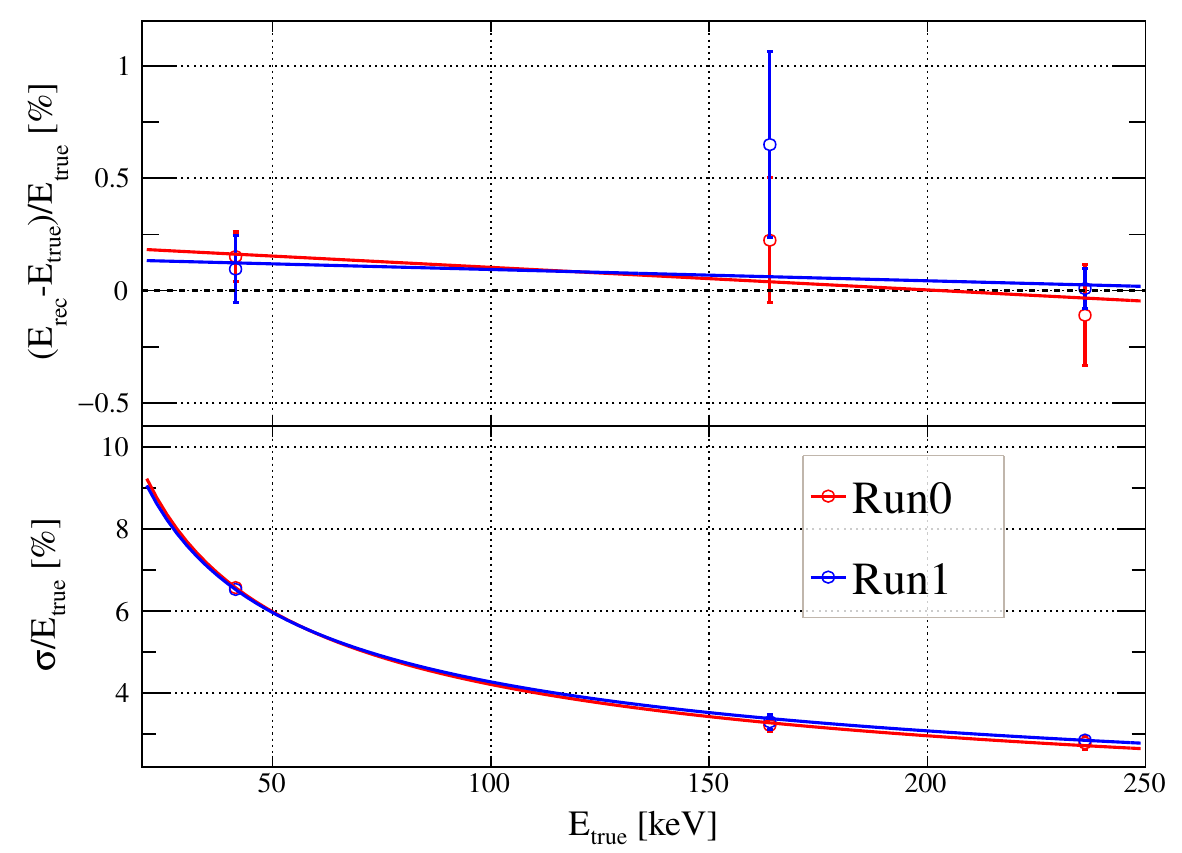}
  \caption{Fractional deviation of energy reconstruction (top) and energy resolution (bottom) for three characteristic gamma peaks in Run0 (red) and Run1 (blue). The error bars reflect both statistical and systematic uncertainties. The red and blue lines represent the corresponding fits to the data.}
  \label{fig:energy}
\end{figure}

The accuracy of energy reconstruction and the energy resolution are verified using the ER peaks mentioned above, as shown in Figure~\ref{fig:energy}.
The deviations of reconstructed energies $E_{\rm rec}$ from their theoretical values $E_{\rm true}$ are smaller than 1\% for the reconstructed energies.
The energy resolution is defined by calculating the ratio of the standard deviation $\sigma$ to the expected energy $E_{\rm true}$, obtained from Gaussian fits to the reconstructed spectra.
We model the energy deviation and energy resolution as functions of energy as 
\begin{equation}\label{eq:linearity}
 (E_{\rm rec}-E_{\rm true})/E_{\rm true} = a_{0}+b_{0}E_{\rm true}.
\end{equation}
and
\begin{equation}\label{eq:resolution}
    \frac{\sigma}{E_{\rm true}} = \frac{a_{\rm res}}{\sqrt{E_{\rm true}}} + b_{\rm res}
\end{equation}
The four parameters ($a_{0}$, $b_{0}$, $a_{\rm res}$, $b_{\rm res}$) within the formula will be treated as constrained variables, which will be incorporated into the final likelihood function fit process, as detailed in Sec.~\ref{subsec.FitMethod}.

\subsection{Signal model}\label{subsec.sig}
The signals of 2$\nu$DEC are from X-rays and/or Auger electrons. 
The energy carried away by the emitted neutrinos, as shown in Eq.~\ref{eq:DEC}, cannot be measured.
The recoil energy of nuclei is negligible.
The energy deposition and the capture fraction from Ref.~\cite{Nitescu:2024ppf}, as summarized in Table~\ref{tab:capture_fraction}, are used for the signal model. An alternative set of energy depositions in Ref.~\cite{XENON:2022evz}, which considers the ionization threshold in xenon, is used as a source of the systematic uncertainty in Sec.~\ref{sec.Result}.

We consider the 2$\nu$DEC signal as five mono-energetic peaks in the ROI, ranging from the $KK$ to $KO$ transitions, with corresponding energies and fractions listed in Table~\ref{tab:capture_fraction}.
The overall signal spectrum is represented as a weighted sum of multiple Gaussian peaks with their resolutions determined by Eq.~\ref{eq:resolution}.
The energy resolutions corresponding to the $KK$ and $KL$ captures of $^{124}$Xe 2$\nu$DEC are (5.3 $\pm$ 0.1)$\%$ and (6.9 $\pm$ 0.1)$\%$, respectively.

\begin{table}[tb]
    \centering
    \caption{Summary of energies and capture fractions of $^{124}$Xe 2$\nu$DEC~\cite{Nitescu:2024ppf} used in this analysis.}
    \begin{tabular}{c|c|c}
        \hline
        \hline
        Decay mode & Energy (keV) & Fraction (\%)\\
        \hline
        $KK$ & 64.62 & 74.14 \\
        $KL$ & 37.05 & 18.80 \\
        $KM$ & 32.98 & 3.84  \\
        $KN$ & 32.11 & 0.84  \\
        $KO$ & 31.93 & 0.13  \\
        \hline
        \hline
    \end{tabular}
    \label{tab:capture_fraction}
\end{table}

\subsection{Background model}\label{subsec.bkg}
There are two main categories of background in this analysis.
Linear or continuous background spectra are extended from the dark matter search region into the ROI, including contributions from $^{214}$Pb, $^{212}$Pb, $^{85}$Kr, material radioisotopes, $^{136}$Xe, and solar neutrinos.
Characteristic energy spectra specifically within the ROI are from cosmogenic and neutron-activated isotopes, including $^{127}$Xe, $^{133}$Xe, and $^{125}$I.
A two-dimensional background model that incorporates energy and temporal information is employed to enhance the identification efficiency of the $^{124}$Xe 2$\nu$DEC signal.
The various background types, their energy and temporal signatures, and the estimated event counts are summarized in Table~\ref{tab:bkg_pdf_count_budget} and are discussed in detail below.

\begin{table}[tbp]
    \centering
    \caption{Summary of the background components, their spectrum and evolution signature, and corresponding expected and best-fit event counts in different data sets.}
    \begin{tabular}{c|c|c|c|c|c}
        \hline
        \hline
        Backgrounds & Spectrum & Time evolution & Dataset & Expected & Best-fit\\
        \hline
        $^{214}$Pb  & Linear & $^{222}$Rn $\alpha$ & Run0 & 748$\pm$59  & 732$\pm$51\\
                    & &                            & Run1 & 1641$\pm$130& 1768$\pm$104\\        
        $^{212}$Pb  & Linear & Constant            & Run0 & 130$\pm$35  & 130$\pm$32\\
                    & &                            & Run1 & 188$\pm$50  & 193$\pm$47\\
        $^{85}$Kr   & Linear & Constant            & Run0 & 209$\pm$108 & 218$\pm$63\\
                    & &                            & Run1 & 658$\pm$196 & 741$\pm$115\\
        Material    & Linear & Constant            & Run0 & 146$\pm$6   & 145$\pm$6\\
                    & &                            & Run1 & 285$\pm$11  & 284$\pm$12\\
        $^{136}$Xe  & Linear & Constant            & Run0 & 245$\pm$11  & 243$\pm$11\\
                    & &                            & Run1 & 437$\pm$20  & 436$\pm$20\\
        Solar $\nu$ & Continuous & Constant        & Run0 & 85$\pm$9    & 85$\pm$8\\
                    & &                            & Run1 & 153$\pm$15  & 153$\pm$15\\
        $^{127}$Xe  & Gaussian @33\,keV & Decay    & Run0 & 43$\pm$1    & 43$\pm$1\\
        $^{133}$Xe  & Tail into 75\,keV & Decay    & Run0$_{a}$ & 0.18$\pm$0.06 & 0.17$\pm$0.04\\
                    & &                            & Run0$_{b}$ & 0.00$\pm$0.03 & 0.00$\pm$0.01\\
                    & &                            & Run0$_{c}$ & 0.4$\pm$0.1   & 0.4$\pm$0.1\\
                    & &                            & Run0$_{d}$ & 20$\pm$2      & 20$\pm$2\\
        Rapid $^{125}$I & Multi-Gaussian & Decay   & Run0$_{b}$ & - & 7$\pm$10\\
                        & &                        & Run0$_{c}$ & - & 10$\pm$8\\
                        & &                        & Run0$_{d}$ & - & 56$\pm$23\\
        Slow $^{125}$I  & Multi-Gaussian & Decay   & Run0$_{a}$ & - & 14$\pm$7\\
                        & &                        & Run0$_{b}$ & - & 13$\pm$7\\
                        & &                        & Run0$_{c}$ & - & 23$\pm$12\\
                        & &                        & Run0$_{d}$ & - & 92$\pm$28\\
                        & &                        & Run1       & - & 28$\pm$42\\
        \hline 
        \hline
    \end{tabular}
    \label{tab:bkg_pdf_count_budget}
\end{table}

\subsubsection{General backgrounds}\label{subsubsec.general}
The $\beta$ decay of $^{214}$Pb, a progeny of $^{222}$Rn, constitutes one of the most significant background contributions.
A depletion effect has been observed~\cite{Ma:2020kll}, where the decay rate of $^{214}$Pb decreases relative to that of $^{222}$Rn.
The average $^{214}$Pb decay rate of $4.5 \pm 0.2\,\mu$Bq/kg in Run0 is directly measured by fitting its broad $\beta$ spectrum at high energy~\cite{PandaX:2023ggs}, yielding a depletion factor of 63\%.
The same factor is applied to Run1, resulting in an average rate of $5.5 \pm 0.2\,\mu$Bq/kg based on the measured $^{222}$Rn $\alpha$ rate.
The $^{214}$Pb contribution within the ROI is estimated by evaluating the ratio of events in the ROI to the full spectrum from a dedicated $^{222}$Rn calibration run and a high-precision theoretical calculation~\cite{Haselschwardt:2020iey}.
The levels of $^{222}$Rn and $^{214}$Pb varied throughout both Run0 and Run1 due to online distillation and xenon circulation adjustments,
and the temporal evolution of $^{214}$Pb is tracked through the measured $^{222}$Rn $\alpha$ rate.

The background contribution from the $\beta$ decay of $^{212}$Pb, a daughter nucleus in the $^{220}$Rn decay chain, is estimated using procedures similar to those for $^{214}$Pb.
The $^{212}$Pb activity is determined from the $^{212}$Po $\alpha$ rate and the $^{212}$Pb/$^{212}$Po ratio, which is derived from the high-energy spectrum fit in Ref.~\cite{PandaX:2023ggs}.
The resulting activity is $0.30 \pm 0.08$ ($0.24 \pm 0.06$)\,$\mu$Bq/kg in Run0 (Run1).
Since no temporal variations in $^{212}$Po $\alpha$ rates are observed beyond statistical fluctuations, the above values are taken as stable activities of $^{212}$Pb in Run0 and Run1.

The background from $^{85}$Kr $\beta$ decay is evaluated  by the correlated $\beta$-$\gamma$ emissions from the $^{85\rm m}$Rb.
The concentrations of Kr in Run0 and Run1 are estimated to be $0.52 \pm 0.27$ and $0.94 \pm 0.28$ parts per trillion, respectively, assuming a $^{85}$Kr abundance of $2 \times 10^{-11}$~\cite{Collon:2004xs}.
The $^{85}$Kr background rate is assumed to remain constant over both Run0 and Run1.

The radioactivity in the detector materials primarily originates from $^{60}$Co, $^{40}$K, $^{232}$Th, and $^{238}$U in the PMTs and detector vessels.
The radioactivity is measured with a high-purity germanium detector~\cite{PandaX-4T:2021lbm} and subsequent wide energy spectrum fits, as detailed in Ref.~\cite{PandaX-4T:2022kwg}.
Their background contributions within the ROI are evaluated using the Geant4-based BambooMC simulation framework~\cite{Chen:2021asx} and combined into a single background component for this analysis.
The contribution from the two-neutrino double $\beta$ decay ($2\nu\beta\beta$) of $^{136}$Xe is constrained using the half-life and isotopic abundance of $^{136}$Xe measured \textit{in situ} by PandaX-4T~\cite{PandaX-4T:2022kwg}.
The background from solar $pp$ and $^7$Be neutrinos elastically scattering on electrons is estimated with the event rate and continuous spectrum calculated in Ref.~\cite{Chen:2016eab}.
A 10\% uncertainty in the solar neutrino flux, based on the Borexino measurement~\cite{BOREXINO:2018ohr}, is included in this analysis.
The background rates from material radioactivity, $^{136}$Xe $2\nu\beta\beta$, and solar neutrinos are assumed to remain unchanged in our detector, as detector operation campaigns do not influence them.
Except for the continuous spectrum of the solar neutrino background, the spectra of all other backgrounds described above are generated from the BambooMC simulation and parameterized with a linear function in the ROI.

\subsubsection{Cosmogenic and neutron activation backgrounds}\label{subsubsec.activation}
During Run0, a bottle of above-ground xenon was injected into the detector, represented by the yellow shaded region in Figure~\ref{fig:exposure_operation}, leading to a significant increase in cosmogenic $^{127}$Xe activity.
The decay of $^{127}$Xe via electron capture (EC) contributes to the background at 33.2\,keV, with a half-life of 36.4 days~\cite{LNHB:Xe127}.
The amount of 33.2\,keV events can be constrained by the ratio of this peak with a mono-energetic peak at 408\,keV, where 375\,keV $\gamma$-rays from the daughter $^{127}$I nucleus deposit the full energy in the TPC.
The ratio is simulated with BambooMC.
The 408\,keV events are also utilized to track the evolution of the $^{127}$Xe background throughout Run0.
The total $^{127}$Xe background in Run0 is estimated to be $43 \pm 1$ events.
The $^{127}$Xe contribution in Run1 is negligible due to the long interruption of approximately 6 months after the end of Run0.

$^{133}$Xe and $^{125}$Xe were also introduced into PandaX-4T via neutron capture during the cosmic exposure and neutron source calibration runs.
Different types of neutron sources, including $^{241}$Am-Be (AmBe) and $^{238}$Pu-C(PuC), were used in calibrations before, during, and after Run0, as shown in Figure~\ref{fig:exposure_operation}.
Among them, the PuC source has the strongest neutron flux.
Specifically, the xenon injection, as well as the AmBe and PuC calibrations during Run0, introduced additional $^{133}$Xe and $^{125}$Xe in Run0.
Therefore, Run0 is correspondingly divided into four subsets, labeled Run0$_{a}$, Run0$_{b}$, Run0$_{c}$, and Run0$_{d}$ for convenience, as illustrated in Figure~\ref{fig:exposure_operation}.
Their respective live times are 15.5, 14.1, 27.1, and 38.1 days.
Additionally, the intensive neutron calibration conducted after Run0 may introduce residual backgrounds that persist into Run1.
No neutron calibration run was conducted during the Run1 data-taking period.

$^{133}$Xe is a critical background near the 75\,keV right edge of the ROI in this search.
It undergoes $\beta$ decay to $^{133}$Cs with a half-life of 5.2\,days and transitions to the 80.1\,keV excited state with a capture fraction of 98.5\%~\cite{LNHB:Xe133}.
The $\beta$-$\gamma$ coincidence events starting at 80.1\,keV may leak into the ROI due to the energy smearing.
By selecting events in the 75$-$120\,keV range for an energy-time two-dimensional side-band fit using the same background model as the final fit (See Sec.~\ref{subsec.FitMethod}), we can derive an estimation and constrain the level of $^{133}$Xe in the ROI.
The $^{133}$Xe backgrounds in four subsets, Run0$_{a}-$Run0$_{d}$, are determined to be $0.18 \pm 0.06$, $0.00 \pm 0.03$, $0.4 \pm 0.1$, and $20 \pm 2$ events, respectively.
The contribution of $^{133}$Xe is not considered in Run1 due to its short half-life and the relatively long interval between Run0 and Run1.

$^{125}$Xe, or the subsequent $^{125}$I, is another key background in the search for $^{124}$Xe 2$\nu$DEC signals.
$^{125}$Xe undergoes EC with a half-life of 16.9 hours, decaying into the relatively long-lived $^{125}$I, which has a half-life of 59.4 days~\cite{LNHB:I125}.
The $^{125}$Xe EC decay generates 33.2\,keV depositions, most of which are associated with 243.4\,keV or 188.4\,keV $\gamma$-rays, and the estimated background within the ROI is negligible; thus it is not included in the background model.
$^{125}$I, however, decays via EC to the 35.5\,keV excited state of $^{125}$Te, with $\sim{}$80$\%$ of the decays capturing electrons on the K-shell, $\sim{}$16$\%$ on the L-shell, and $\sim{}$3.5$\%$ on the M-shell~\cite{i125k:ToRad}.
The total energy depositions for these EC processes are 67.3\,keV, 40.4\,keV, and 36.5\,keV, respectively~\cite{Deslattes:2003zz, xray:LBL}.
These energy peaks closely coincide with those of the $^{124}$Xe 2$\nu$DEC, making $^{125}$I the most significant background component in this analysis.

$^{125}$I potentially existed in the Run0 subsets and Run1, predominantly in Run0$_{d}$ after the PuC calibration.
The temporal evolution of $^{125}$I inside the FV following neutron activation can be modeled as

\begin{equation}
 N_{\rm {^{125}I}}(t) = -\kappa_1 e^{-(t-t_0)/\tau_{\rm {^{125}Xe}}} + \kappa_2 e^{-(t-t_0)/\tau_{\rm eff}} + \kappa_3 e^{-(t-t_0)/\tau_{\rm {^{125}I}}}.
 \label{eq:I125model}
\end{equation}

The first term describes the production of $^{125}$I from the decay of $^{125}$Xe according to its radioactive lifetime $\tau_{\rm {^{125}Xe}}$.
The second term accounts for both the physical decay of $^{125}$I and its continuous removal from the FV via the xenon circulation and purification system, with an effective lifetime $\tau_{\rm eff}$, which is expected to be much shorter than the physical decay lifetime $\tau_{\rm {^{125}I}}$.
Due to the stable circulation flow, the removal of $^{125}$I is consistently described by a common $\tau_{\rm eff}$ throughout Run0.
The third term represents the potential contribution of $^{125}$I from the circulation-inaccessible dead zones, which diffuses into the FV and decays with its physical lifetime.
A consistent phenomenon was also observed during the removal of tritium in PandaX-II~\cite{PandaX-II:2017hlx}.
For clarity, we refer to this component as ``slow $^{125}$I'', and the others as ``rapid $^{125}$I''.

The rapid $^{125}$I component appeared only in the subsets of Run0$_{b}$, Run0$_{c}$, and Run0$_{d}$ due to its short $\tau_{\rm eff}$.
The contributions from rapid $^{125}$I in Run0$_{a}$ and Run1 are negligible since the related neutron calibration occurred several months prior to these datasets.
Comparatively, the slow $^{125}$I component persists throughout the entire PandaX-4T data collection period due to its long lifetime.
These contributions are floated in the Run0 subsets and Run1 in the final likelihood fit.

\subsection{Fit method and half-life calculation}\label{subsec.FitMethod}
We use the unbinned maximum likelihood method to incorporate the temporal evolution of various backgrounds in the analysis.
The likelihood function is constructed as

\begin{multline}
\label{eq:likelihood}
\mathcal{L} = \left\{\prod_{n=\rm 0,1} \left[
 {\rm Poisson}(\mathcal{N}_{\text{obs}}^n|\mathcal{N}_{\text{fit}}^n)
    \times
    \prod_{i=1}^{\mathcal{N}_{\text{obs}}^n} (l_{s}^{n,i}+\sum_b {l_{b}^{n,i}})\right]\right\}\\
\times\left[\prod_{b,n}G(N_{b,\text{fit}}^{n}, N_b^{n}, \sigma_b^{n}) \mathcal{G}(\Vec{p}, \Vec{\mu_p}, \Vec{\sigma_p})\right],
\end{multline}
with
\begin{align}
 &\mathcal{N}_{\text{fit}}^{n}=N_{s,\text{fit}}^n + \sum_bN_{b,\text{fit}}^n,\\
 \label{eq:frac_likelihood_sig}  &l_{s}^{n,i}=\frac{N_{s,\text{fit}}^nP_{\rm s}^n(E^i,t^i;\Vec{p})}{\mathcal{N}_{\text{fit}}^{n}},\\
 \label{eq:frac_likelihood_bkg}  &l_{b}^{n,i}=\frac{N_{b,\text{fit}}^nP_{b}^n(E^i,t^i;\Vec{p})}{\mathcal{N}_{\text{fit}}^{n}}.
\end{align}

To distinguish between the two experimental phases, we denote the runs with the variable $n$, where $n=0$ corresponds to Run0 and $n=1$ corresponds to Run1.
For each run, the number of observed events is $\mathcal{N}_{\text{obs}}^n$, and that of total fitted events is $\mathcal{N}_{\text{fit}}^n$.
The number of fitted background $N_{b,\text{fit}}^{n}$ in each data set is constrained by the corresponding expected number $N_b^{n}$ and uncertainty $\sigma_b^{n}$ with a Gaussian penalty function $G(N_{b,\text{fit}}^{n}, N_b^{n}, \sigma_b^{n})$.
For backgrounds including $^{214}$Pb, $^{212}$Pb, $^{85}$Kr, material radioactivity, $^{136}$Xe, and solar $\nu$, two independent nuisance parameters of $N_{b,\text{fit}}^{n}$ are assumed for Run0 and Run1 respectively to reflect the potential changes in run conditions.
The $^{127}$Xe background has a single nuisance parameter in Run0, while the $^{133}$Xe background is modeled with four independent parameters of $N_{b,\text{fit}}^n$ in the four subsets of Run0.
Both $^{127}$Xe and $^{133}$Xe backgrounds are considered negligible in Run1.
The $^{124}$Xe signal, along with the rapid and slow components of the $^{125}$I background, are left to float in the corresponding data sets and are characterized through their energy and evolution information in the fit.

The PDFs of signal and backgrounds, $P_s^n$ and $P_{b}^n$, have two dimensions ($E$, $t$) and can be decoupled to two independent parts of the energy spectrum and time evolution: $P^n(E, t; \Vec{p}) = P^n(E; \Vec{p_E}) \times P^n(t; \Vec{p_t})$.
$\Vec{p}$ are characteristic parameters describing the energy spectrum ($\Vec{p_E}$) and time evolution ($\Vec{p_t}$) of signal and backgrounds, which include parameters about energy resolution, energy shift, and data selection efficiency, slopes of linear spectra, and the effective lifetime of $^{125}$I.
All parameters except $\tau_{\rm eff}$ are constrained by their nominal values $\Vec{\mu_p}$ and uncertainties $\Vec{\sigma_p}$ through a composite Gaussian penalty function $\mathcal{G}(\Vec{p}, \Vec{\mu_p}, \Vec{\sigma_p})$.
Correlated parameters are constrained by multi-dimensional Gaussian functions, while other parameters are either constrained by one-dimensional Gaussian functions or left as free, as detailed in Table~\ref{tab:vecp}.
\begin{table*}[tbp]
    \centering
    \caption{Summary of the parameters $\vec{p}$.}
    \begin{tabular}{ccc}
    \hline
    \hline
    Parameters &\centering Description & Constraint type\\
    \hline
    $a_{\rm res}$, $b_{\rm res}$ & Energy resolution & 2D Gaussian\\
    \hline
    $a_{0}$, $b_{0}$ & Energy shift & 2D Gaussian\\
    \hline
    $\epsilon_{1}$, $\epsilon_{2}$, $\epsilon_{3}$ & Efficiency curve & 3D Gaussian\\
    \hline
    $k_{b}$ & Slopes of background energy spectra & 1D Gaussian\\
    \hline
    $\tau_{\rm eff}$ & Effective lifetime of rapid $^{125}$I & Free parameter\\
    \hline
    \hline
    \end{tabular}
    \label{tab:vecp}
\end{table*}

For a fitted count of $^{124}$Xe 2$\nu$DEC, $N_{2\nu \rm DEC}$, the half-life is calculated with 
\begin{equation}\label{eq:halflife}
 T_{1/2}^{2\nu \rm DEC} = {\rm ln\,2} \times \frac{N_A\times\eta\times \epsilon \times m \times t}{N_{2\nu \rm DEC}\times M_A},
\end{equation}
where $m \times t$=1.73\, ton$\cdot$yr is the total exposure, $M_A$ = 0.131\,kg/mol is the xenon molar mass, $N_A$ is the Avogadro's constant, $\eta$ = $(10.0 \pm 0.1)\times10^{-4}$ is the isotopic abundance of $^{124}$Xe in PandaX-4T, which is measured by a residual gas analyzer, and $\epsilon=99.4\%$ is the signal efficiency of $^{124}$Xe considering both detection and data quality cuts. 

\section{Results and discussion}\label{sec.Result}
\begin{figure*}[tb]
  \centering
  \includegraphics[width=\columnwidth]{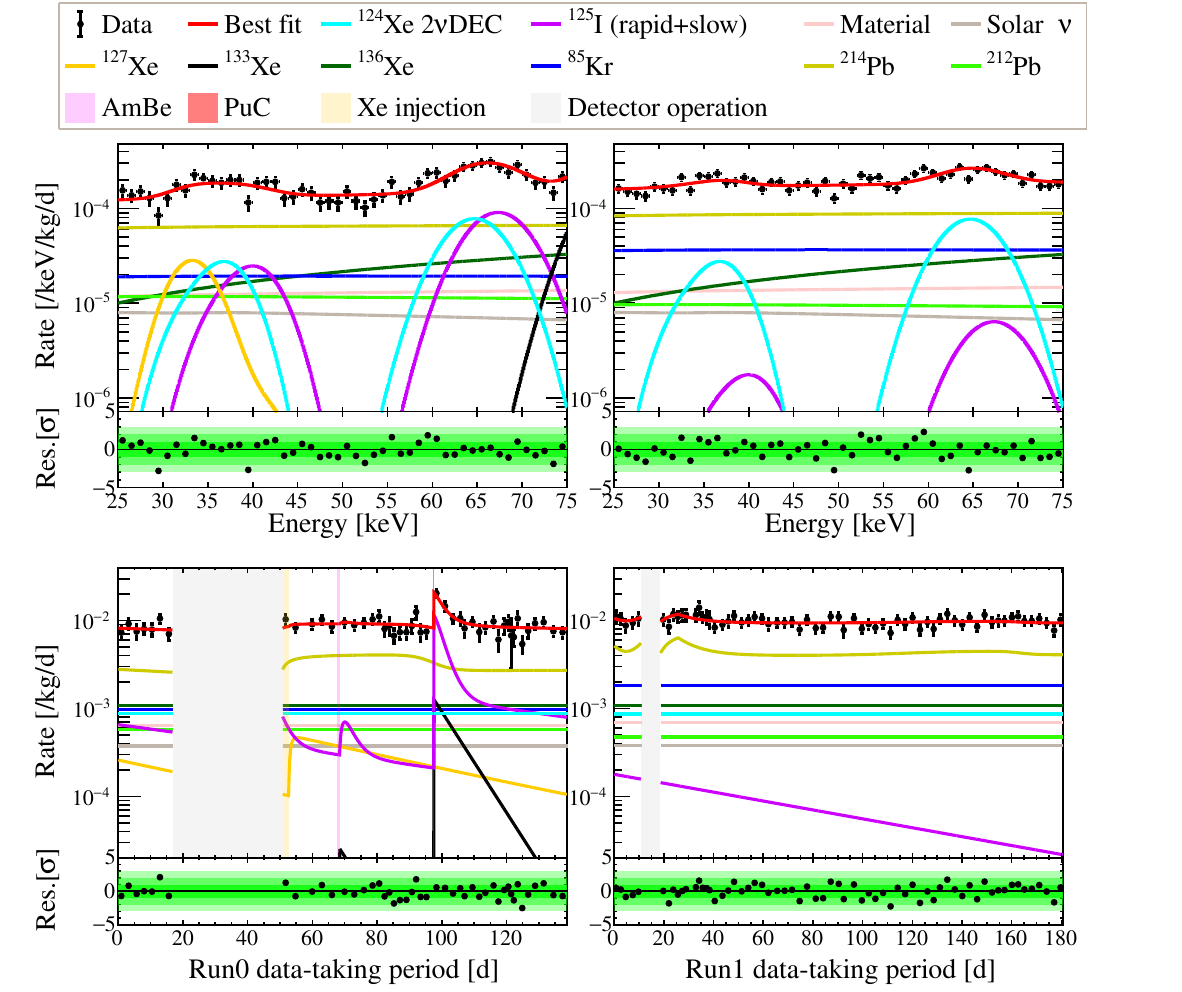}
  \caption{Results of the 2D unbinned likelihood fit to combined Run0 and Run1 data. For better visualization, the results are projected into binned energy spectra (top) and time evolution (bottom) for Run0 (left) and Run1 (right), respectively. The energy spectra in data are divided into 1 keV each bin, and the time evolution is unequally grouped with adjacent data, with 2$\sim$5 days each bin, for clearer display. The red solid line indicates the best-fit signal plus the background model, and the cyan solid line indicates the $^{124}$Xe 2$\nu$DEC signal. Data and fit results are shown in each figure in the top panel, and the corresponding residuals are in the bottom panel.}
  \label{fig:finalfit}
\end{figure*}

The best-fit results, projected onto binned energy spectra and time evolution for both Run0 and Run1, are shown in Figure~\ref{fig:finalfit}.
The fitted curve agrees well with the data, as demonstrated with a goodness-of-fit test on the two-dimensional histogram with 25 by 25 bins. 
The reduced chi-square is 1.06, corresponding to a $p$-value of 0.19.
The best-fit number of $^{124}$Xe 2$\nu$DEC is 549 events, with a statistical uncertainty of 77 events.
The fit results of all backgrounds are listed in Table~\ref{tab:bkg_pdf_count_budget} and compared to their expected values.

For those backgrounds with constraints, the fitted contributions are consistent with their expected values.
Notably, the temporal evolution of the $^{214}$Pb background model, represented by the dark yellow lines in Figure~\ref{fig:finalfit}, aligns well with the observed temporal patterns in both Run0 and Run1 data. 
The fitted $^{214}$Pb activity is in agreement with its expected value within 1\,$\sigma$.

The $^{125}$I peaks around 40\,keV and 70\,keV in the energy spectrum, and the corresponding temporal evolution in Run0 data are well fitted.
The effective lifetime of the rapid $^{125}$I component is fitted as (2.9 $\pm$ 2.7)\,d, which is much smaller than the physical lifetime of $^{125}$I due to the continuous purification of liquid xenon.
The $^{125}$I background was present during both Run0 and Run1, with a notable increase in both the rapid and slow components following the PuC neutron calibration in late Run0 (i.e., Run0$_{d}$).
The phenomenon also agrees with expectation since the PuC neutron source is much stronger than other ones.

The systematic uncertainties consist of eight contributions listed in Table~\ref{tab:uncertainty}.
The first four items are incorporated into the likelihood function with corresponding characteristic parameters $\Vec{p}$ constrained by Gaussian penalties.
The last four items are evaluated independently.

\begin{table*}[tbp]
  \centering
  \caption{Summary of statistical and systematic uncertainties on $^{124}$Xe 2$\nu$DEC half-life.
  }
  \begin{tabular}{cccc}
    \hline
    \hline
    Type & Contribution & Absolute ($10^{21}\,\rm yr$) & Relative (\%)\\
    \hline
    Statistical &                               & 1.51  & 14.6 \\
    \hline
               & Background estimation          & 0.56  & 5.4 \\
               & Energy resolution and linearity& 0.15  & 1.5 \\
               & Selection efficiency           & $<0.01$ & -   \\
               & Fiducial mass                  & 0.05  & 0.5 \\
    Systematic & $^{124}$Xe abundance           & 0.10  & 1.0 \\
               & $^{124}$Xe 2$\nu$DEC model     & 0.19  & 1.8 \\
               & Background models              & $<0.01$ & -   \\
               & Fit range                      & 0.23  & 2.2 \\
    \cline{2-4}
               & Total                          & 0.77  & 7.5 \\
    \hline
    \hline
  \end{tabular}
  \label{tab:uncertainty}
\end{table*}

The uncertainty on the $^{124}$Xe isotopic abundance is obtained as 1.0\% based on the PandaX-4T measurement, taking into consideration both statistical and systematic uncertainties.

The slight difference between the two calculations of energy depositions of different $^{124}$Xe 2$\nu$DEC subshell introduces a variation of 1.8\% in the outcomes, which we account for as a systematic uncertainty of $^{124}$Xe 2$\nu$DEC model.

The uncertainties from the background models in time evolution are also manually estimated, particularly for $^{214}$Pb, whose event rates varied over time.
Different functions, such as switching from exponential to polynomial, are applied to fit the ascending and descending slopes of $^{222}$Rn $\alpha$ decays and used as an alternative model for $^{214}$Pb evolution.
The shape of the background energy spectrum, namely the uncertainty in the slope of the linear function, has also been investigated.
However, these model variations are found to have a negligible impact on the final fit results.

The systematic uncertainty from the fit range is evaluated manually by adjusting the ROI to 24$-$74\,keV and 26$-$76\,keV.
These changes resulted in a small uncertainty of 2.2\% compared to the baseline fit, which also confirms the stability of the fit.

\begin{figure}[tbp]
    \centering
    \includegraphics[width=0.8\columnwidth]{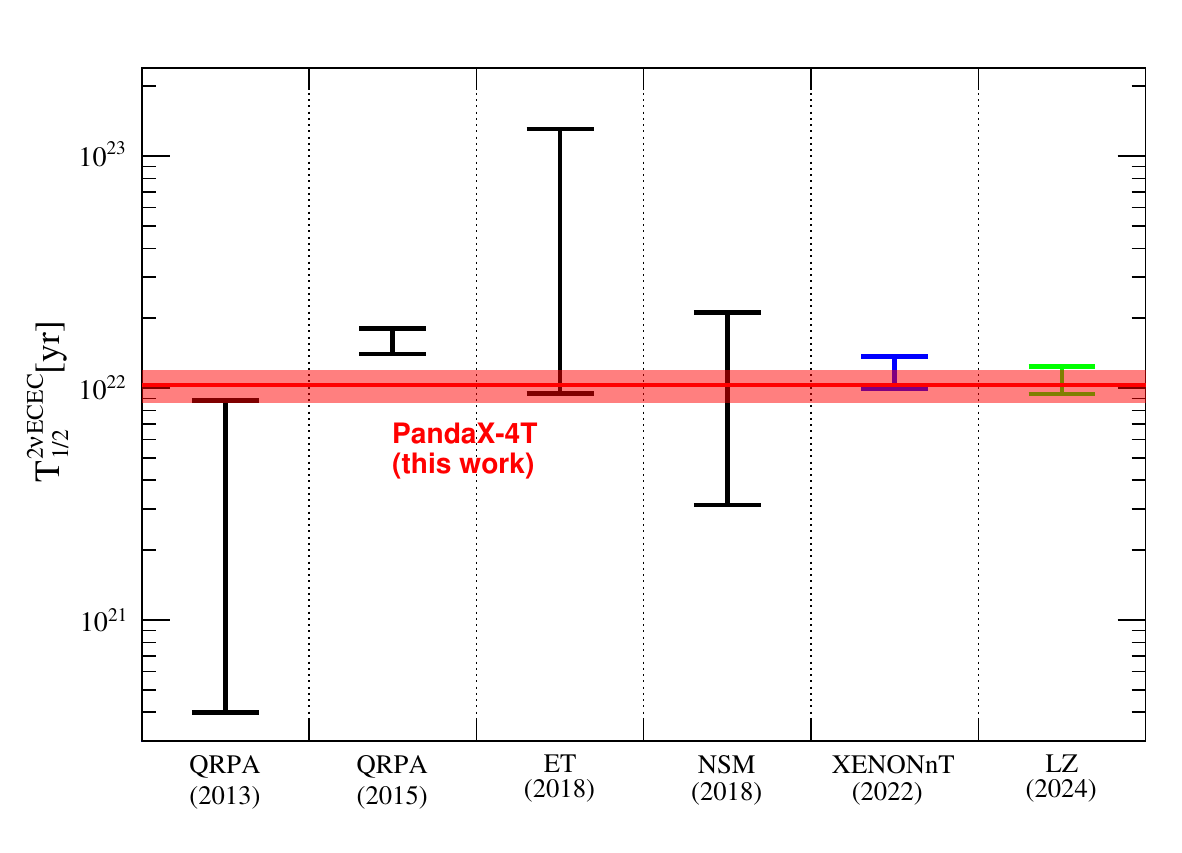}
    \caption{Comparison of the measured half-life with theoretical predictions~\cite{CoelloPerez:2018ghg,Suhonen:2013rca, Pirinen:2015sma} and other experiments~\cite{XENONCollaboration:2022kmb,Aalbers:2024xwo}.}
    \label{fig:nuclear_models}
\end{figure}

The total relative systematic uncertainty is 7.5\%, calculated by summing all the systematic listed in Table~\ref{tab:uncertainty} in quadrature.
The final half-life of $^{124}$Xe 2$\nu$DEC is determined to be 
$(1.03\pm0.15_{\rm stat}\pm0.08_{\rm sys})\times10^{22}$\,yr.
This result is consistent with calculations from the effective theory (ET) and large-scale nuclear structure model (NSM)~\cite{CoelloPerez:2018ghg} within 1\,$\sigma$, and with results from the quasiparticle random-phase approximation (QRPA)~\cite{Suhonen:2013rca, Pirinen:2015sma} within $\sim2\,\sigma$.
It is also in agreement with recent experimental measurements~\cite{XENONCollaboration:2022kmb,Aalbers:2024xwo} within the 1\,$\sigma$ range, as depicted in Figure~\ref{fig:nuclear_models}.

Given the many approximations inherent in the model calculations for the 2$\nu$DEC capture fractions of $^{124}$Xe, experimental measurement of these ratios remains valuable.
In our analysis, we fix the relative values of the $KL/KM/KN/KO$ capture fractions and allow the sum of these three to change along with the $KK$ capture fraction while maintaining the total sum of the five capture fractions constant.
The fitted $KK$ capture fraction of $^{124}$Xe is $(65\pm5)\%$, in reasonable agreement with the calculated value listed in Table~\ref{tab:capture_fraction} at 1.8\,$\sigma$.
It is also consistent with the recent measurement from the LZ collaboration~\cite{Aalbers:2024xwo}.

\section{Conclusion}\label{sec.Conclusion}
In this analysis, we construct an unbinned likelihood using a time-dependent background model to measure the $^{124}$Xe 2$\nu$DEC half-life.
The accuracy and resolution of energy reconstruction are optimized and quantitatively calibrated.
The temporal variation of certain background components is carefully considered.
Using a total of 1.73 tonne$\cdot$year exposure from the commissioning run and the first science run, we obtain the half-life of $^{124}$Xe 2$\nu$DEC as 
$(1.03\pm0.15_{\rm stat}\pm0.08_{\rm sys})\times10^{22}$\,yr. 
The capture fraction of $KK$ capture is measured to be  $(65\pm5)\%$.
Our measurement results are in excellent agreement with the XENON and LZ values~\cite{XENONCollaboration:2022kmb,Aalbers:2024xwo}.
Those measurements will be important inputs for nuclear theory models to calculate the NME of 2$\nu$DEC and 0$\nu$DEC.
The precision results also demonstrate the wide application and full physics potential of large low-background liquid xenon TPCs.

\section*{Acknowledgements}

This project is supported in part by grants from National Science Foundation of China (Nos. 12105052, 12090060, 12090062, 12005131, 11905128, 11925502, U23B2070), grants from China Postdoctoral Science Foundation (Nos. 2021M700859, 2023M744093), a grant from the Ministry of Science and Technology of China (Nos. 2023YFA1606200, 2023YFA1606202), and grants from Office of Science and Technology, Shanghai Municipal Government (grant No. 21TQ1400218, 22JC1410100, 23JC1410200, ZJ2023-ZD-003). We thank for the support by the Fundamental Research Funds for the Central Universities. We also thank the sponsorship from the Chinese Academy of Sciences Center for Excellence in Particle Physics (CCEPP), Hongwen Foundation in Hong Kong, New Cornerstone Science Foundation, Tencent Foundation in China, and Yangyang Development Fund. Finally, we thank the CJPL administration and the Yalong River Hydropower Development Company Ltd. for indispensable logistical support and other help.

\bibliographystyle{apsrev4-1}
\bibliography{refs.bib}


\end{document}